\documentclass[journal]{IEEEtran}

\usepackage{graphicx, caption}
\usepackage{algpseudocode}
\usepackage{algorithm, amsmath}
\usepackage{cite}
\usepackage{booktabs} % For better looking tables
\usepackage{array}    % For table wrapping
\usepackage{hyperref}

\title{Heart Sound Segmentation Using Deep Learning Techniques}
\author{Manas~Madine
\thanks{Manas Madine is with the Department of Computer Science, University of Massachusetts, Amherst, MA, USA (e-mail: mmadine@umass.edu).}%
}

\begin{document}
\maketitle

% Abstract
\begin{abstract}
Heart disease remains a leading cause of mortality worldwide. Auscultation, the process of listening to heart sounds, can be enhanced through computer-aided analysis using Phonocardiogram (PCG) signals. This paper presents a novel approach for heart sound segmentation and classification into S1 (LUB) and S2 (DUB) sounds. We employ FFT-based filtering, dynamic programming for event detection, and a Siamese network for robust classification. Our method demonstrates superior performance on the PASCAL heart sound dataset compared to existing approaches.
\end{abstract}

% Keywords

\IEEEpeerreviewmaketitle

\section{Introduction}
\IEEEPARstart{T}{he} heart diseases, cardiac arrest, heart attacks are dangerously increasing these days and one of the major causes of death in the world is CVD, that is CardioVascular Diseases. In the United States nearly 34.3 percent and in Europe 48 percent of deaths are related to CVDs \cite{Intro_1}. Auscultation is the process of listening to heart sounds using a stethoscope and Computer-aided Auscultation is analyzing the heart sounds recorded via a digital stethoscope or an acoustic stethoscope. Computer-aided Auscultation has proven to be useful in diagnosing CVD using Phonocardiogram (PCG) signals or electrocardiogram (ECG) signals, etc. Analyzing these signals, exploiting and finding different properties from them is very important in CVD detection studies and therefore very useful in public health studies.

As we were using a digital stethoscope that records the heart sound as a Phonocardiogram (PCG) signal, in this paper we worked on the PCG signal. The two widely used publicly available PCG datasets are from the PhysioNet/CinC (2016) and PASCAL (2011) challenges. These two datasets are significantly different, having used different tools for data acquisition, different clinical protocols, and different overlapping noise that affects the signal qualities \cite{Intro_2}. In this paper, we work on heart sound segmentation and classification of the segmented part of the signal to either S1 (LUB) or S2 (DUB). As the PASCAL dataset has the segmentation times of the heart sounds, we used the PASCAL (Pattern Analysis, Statistical Modelling and Computational Learning) Classifying Heart Sounds Challenge dataset for this paper.

\section{RELATED WORK}
The Heart Sound segmentation can be divided into PreProcessing, event detection, event filtering, and classification.
PreProcessing: PCGs are filtered using different filters. A low-pass Butterworth filter with a 25 Hz cutoff to limit low-frequency noise was used by \cite{RP1}. The majority of the frequency content in S1 and S2 sounds is below 150 Hz, usually with a peak around 50 Hz, so a Bandpass filter was applied to create a signal with 30–60 Hz pass-band by \cite{RP2}. The signal is filtered with a fourth-order Butterworth bandpass filter with cut-off frequencies at 25 Hz and 400 Hz \cite{RP3, RP4, RP5}.
According to \cite{r1}, many methods have been employed previously for Event Detection, such as Envelope-based Methods, feature-based methods, machine learning methods, HMM methods, Logistic regression hidden semi-Markov model (LR-HSMM), and using ECG's Reference \cite{r2}. Among these, the envelope-based methods are popular.
Classification: The detected events are classified as S1 or S2 using different methods. Different features, including frequency domain features \cite{RC1} and wavelets \cite{RC2}, are extracted, and using those, events are classified into their respective states. In \cite{RC3}, the authors propose an event detection-based approach where they utilize RNNs to classify the PCG signal, generating a prediction for each frame in the PCG. Their experimental evaluation suggests that RNN-based temporal modeling is more robust when encountering noisy observations. In \cite{RC4}, instead of classifying each frame like \cite{RC3}, they classified windows using Bidirectional LSTM with attention.
Generic filtering methods use a range of frequency (lower, upper) and filter the signal empirically. We employed FFT-based filtering along with Butterworth filter. FFT filtering is with a unique range for each PCG signal. The methods for classifying each frame and window like \cite{RC3, RC4} to find S1 and S2 will completely rely on the performance of the model. As the S1-S1 duration is important for us, we used onset-based event detection, and for classifying each onset event, we used a Siamese network as it performs well despite the lack of good annotated data.

\section{METHOD}
\subsection{\textbf{Pre-Processing}} Initially, the PCG signal is passed through a fifth-order Butterworth Bandpass filter with cut-off frequencies 20 and 250, which we chose empirically (best suited for our dataset).
According to \cite{Model1}, the frequency spectra were seen to contain peaks in the low-frequency range (10–50 Hz) and the medium-frequency range (50–140 Hz). The S1, S2 peak frequencies are probably related to the elastic properties of the heart muscles and the dynamic events causing those sounds. We observed similar results in the Fast Fourier Transform (FFT) of PCG signals of our dataset. We observed a two-peak signature graph, as shown in figure-1, with peak 1 in the 19-50 Hz range and peak 2 in the 50-100 Hz range.

\begin{figure}[htp]
    \centering
    \includegraphics[width=5cm]{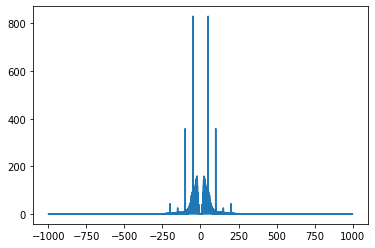}
    \caption{FFT of PCG}
    \label{fig:FFT of PCG}
\end{figure}

Transformation of the signal from the time domain to the frequency domain is the Fourier transform. A Fourier transform to a continuous time signal \( x(t) \) is mathematically written as: 
\begin{equation}
X(w) = \int x(t)e^{jwt} dt 
\end{equation}

we extract the top n dominant frequencies in FFT \( f_{1.....n} \)

\begin{equation}
f_i = argmax_i(X(w))
\end{equation}

we now filter using these dominant frequencies. Instead of using exact frequency, we generate a small range using \(\delta\) before that we sort all the \( f_{1.....n} \) to obtain \( i_{1....n} \), and now we eliminate all the non-dominant frequencies by setting their FFT to zero:

\begin{align*}
X[-\infty.....{i_1 - \delta} ] = 0 \\ 
X[i_1+\delta ..... i_2 - \delta] = 0 \\
. \\ 
. \\
. \\
. \\
X[i_n + \delta ..... +\infty] = 0 
\end{align*}

Now we reconstruct the signal in the time domain using Inverse Fast Fourier Transform (IFFT)

If \( X(w) \) is known, it can be used to obtain value \( x(y) \) using the
inverse Fourier transformation equation. 

\begin{equation}
X(t) = \frac{1}{2\pi} \int x(w)e^{jwt} dw 
\end{equation}

n and \(\delta\) are chosen empirically; we used n = 2 and \(\delta\) = 10.

\begin{figure}[htp]
    \centering
    \includegraphics[width=5cm]{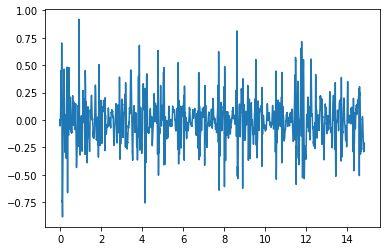}
    \caption{FFT of PCG}
    \label{fig:PCG_signal}
\end{figure}
\begin{figure}[htp]
    \centering
    \includegraphics[width=5cm]{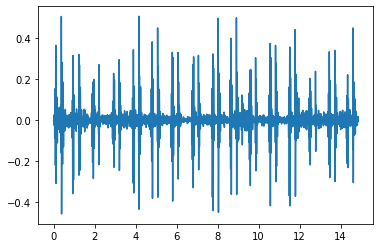}
    \caption{Butter-worth filtered PCG}
    \label{fig:PCG_signal}
\end{figure}
\begin{figure}[htp]
    \centering
    \includegraphics[width=5cm]{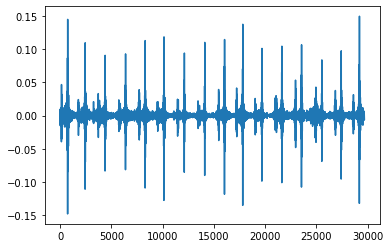}
    \caption{FFT filtered PCG}
    \label{fig:PCG_signal}
\end{figure}

\subsection{\textbf{Event Detection}}
Now the signal is free from most of the noises, we can process the PCG signal to detect events.  
\subsubsection{\textbf {Using Energy Envelop}} Onset Detection, Beat detection, tempo detection have always been extensively studied in Digital Signal Processing, we can employ those techniques here as well. The onset event detection has many steps which include computing spectral flux onset strength envelope, onset events detection by picking peaks in an onset strength envelope and to reduce the overall segmentation error, we backtrack from the peak to locate the peaks corresponding to energy minimum.

\textbf{Onset Strength envelope:}
Onset detection is the process of finding the starting points of all musically relevant events in an audio performance. Onset energy finding is done by summing the squared magnitude energy of the FFT across 40 ranges of frequencies defined by the mel scale with a triangular shape to emphasize a different center frequency.

\begin{equation}
Onset(t) = mean_f max(0, S[f, t] - ref[f, t - lag])
\end{equation}

where \( S \) will be the log-power Mel spectrogram and \( ref \) is \( S \) after local max filtering along the frequency axis \cite{librosa}.

\textbf{Onset Detection:}
Locate note onset events by picking peaks in an onset strength envelope.

\textbf{Backtracking:}
To reduce the overall segmentation error, we backtrack from the detected peak to its corresponding local minimum of its energy envelope. This way we roll back the timing of detected onsets from a peak amplitude to the preceding minimum or the exact starting point of the event.

\subsubsection{\textbf{DL methods for onset Detection}}

\subsection{\textbf{Event Selection}}
As one of our intentions is to find out the heart beat that is the number of times the heart is beating per minute or to put it in other words number of S1, S2 cycles per minute. So from the detected onsets we have to select the onsets which actually correspond to the beginning of the S1 sound and end of S2 sound so that the actual S1, S2 duration is known. 
\subsubsection{\textbf{Dynamic Programming Formulation to find S1 and S2}}
The goal of this algorithm is to select the onset events which are corresponding to beginning of S1 and ending of S1, beginning of S2 and ending of S2. The objective function should take into consideration how good a particular onset event is and does it effectively maintain proper tempo with preceding onset events.
The goodness of an onset event can be found with its corresponding Onset Energy Envelope strength, to check if it maintains a proper tempo or not we had to formulate functions, the main property of that function is that if the present onset is at a proper distance (Threshold distance) to its preceding onset it has to give a better value output compared to if the distance is less that threshold it gives a big penalty output, whereas if the distance is more than the threshold we penalize less. So the perfect function for this is 

\begin{figure}[htp]
    \centering
    \includegraphics[width=5cm]{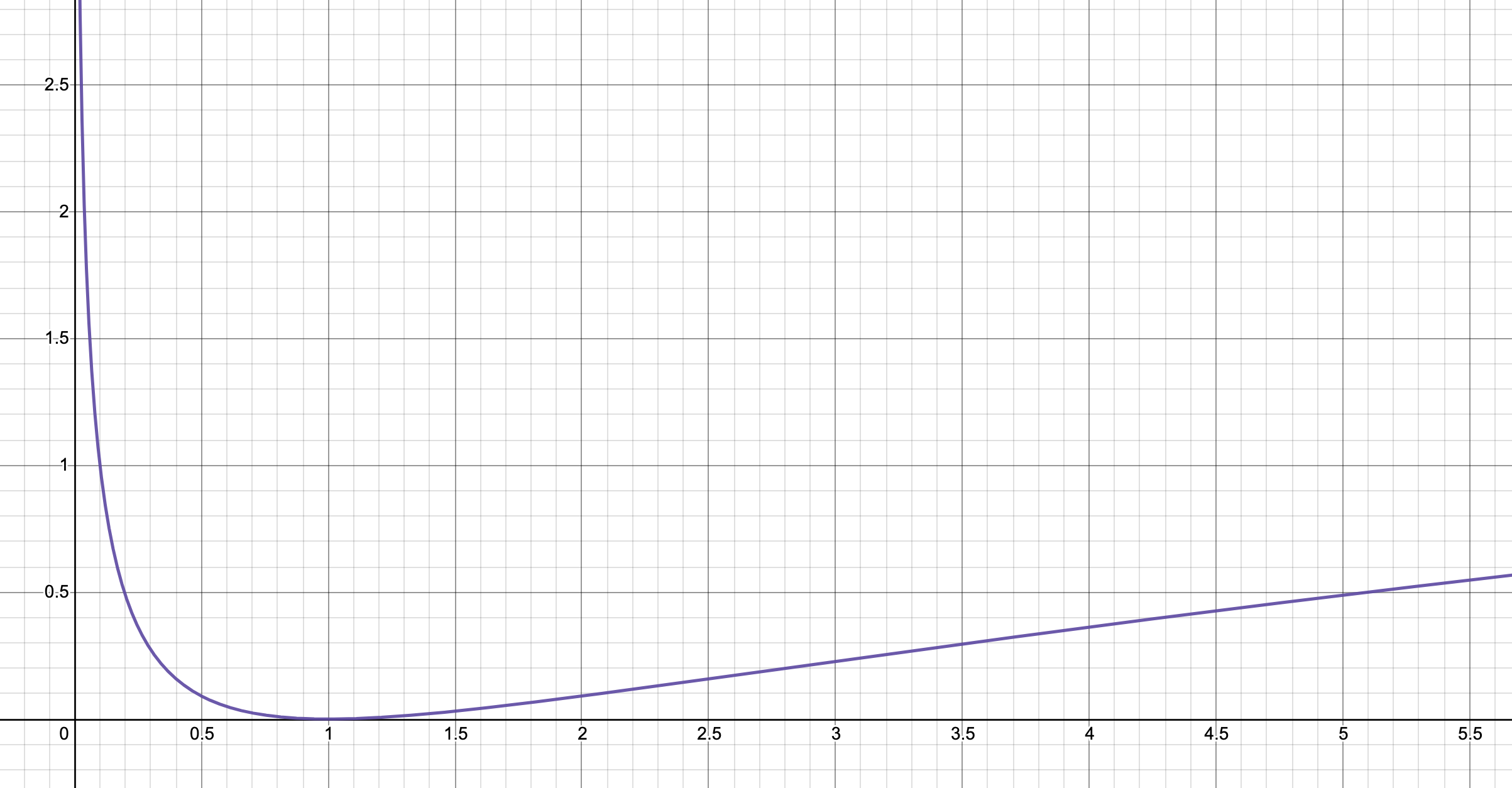}
    \caption{\( \lvert F(\Delta t, \tau) \rvert \)}
    \label{fig:PCG_signal}
\end{figure}

\begin{equation}
F(\Delta t , \tau) = - \alpha(\log(\frac{\Delta t}{\tau}))^2
\end{equation}

where \(\tau\) is the threshold.
Using this function we use two Thresholds [\(\tau_1 , \tau_2\)], \(\tau_1\) is used to define S1-S2 distance threshold, whereas \(\tau_2\) is used to define the S2-S1 distance threshold.

\begin{algorithm}
\caption{Dynamic Programming algorithm}\label{alg:cap}
\begin{algorithmic}
\Require $onsets : t_0,t_2,...t_n$
\Require $n \geq 0$
\State $DP(0 \dots n) \gets o\_env(0 \dots n) $
\State $N \gets n$

\Function{F}{$\Delta t, \tau $}
 \State return $-\alpha(\log(\frac{\Delta t}{\tau}))^2$
\EndFunction

\For{$i \in \{1,\dots,N\}$}
  \For{$j \in \{i-1,\dots,0\}$}
    \State $\Delta t \gets (t_i - t_j)$
    \If {$\Delta t < \delta$ }
        \State $Pre\_Dp(t_i) \gets DP(t_i)$
        \State \begin{align*}
        DP(t_i) = o\_env(t_i) +  max(DP(t_i)  ,DP(t_j) + \\  F(t_i-t_j , \tau_1) , DP(t_j) + F(t_i-t_j , \tau_2))
        \end{align*}
        \If {$Pre\_Dp(t_i) < DP(t_i) $ } 
            \State $Parent_i \gets j $
            \If{$F(t_i-t_j , \tau_1) < F(t_i-t_j , \tau_2)$}
                \State $State_i \gets S2$
            \Else 
                \State $State_i \gets S1$
            \EndIf
        \EndIf
    \Else 
    \State \texttt{break }
    \EndIf     

  \EndFor
\EndFor
\end{algorithmic}
\end{algorithm}

Here, $\alpha$ is the control factor on how much preference is to be given to $o\_env$ strength compared to the penalizing function. We initialize all the DP values of all the onsets to their corresponding onset strength envelope. Then for each onset $t_i$ we iterate from $t_i-1$ to $0$, but to reduce the computational cost we find out $\Delta t$, which is the difference ($t_i$ - $t_j$) and if the difference is more than threshold ($\delta$) we stop the iteration. This is because the S1 - S2 and S2 - S1 distance has a definite maximum limit. We need not search for corresponding S1 or S2 of the present onset beyond threshold ($\delta$).

$DP(t_i)$ is the summation of the onset envelope strength and the maximum of the three values. The first one is $DP(t_j)$ which is initialized to its corresponding onset envelope strength. The second one is the summation of the present DP value and the penalizing function value considering the previous onset to be S1 onset. Then we use the $\tau_1$ which represents the threshold of S1-S2 distance. The third one is the summation of the present DP value and the penalizing function value considering the previous onset to be S2 onset. Then we use the $\tau_2$ which represents the threshold of S2-S1 distance.

If the DP of the present onset changes, it will be greater than the previous DP value ($Pre_DP$). That means the $j$th onset is the parent of the current onset, so we store the parent value as $j$. To remember the state of the current onset, we use the penalizing function. If  $F(t_i-t_j , \tau_1) < F(t_i-t_j , \tau_2)$, that means the $j$th onset and the $i$th onset are having the least penalty with $\tau_1$ which is the threshold for S1-S2. So it means the $j$th (previous onset) is the S1 and the current onset is S2. For each iteration whenever the current DP value changes we update the state and will have the final state corresponding to the maximum DP value in the entire iteration.

We backtrack from the $N$th onset to the onset having time just greater than, $t_n - \delta$ and pick the onset having the peak/maximum value of the DP. This means that the onset corresponding to the maximum DP value is the end of either the S1 or S2 and is the last detected heart sound. As we stored whether the onset is S1 or S2, we now backtrack to the parent of this onset which is nothing but the onset which gave the minimum value of the penalizing function.
\begin{figure*}
    \includegraphics[width=\textwidth]{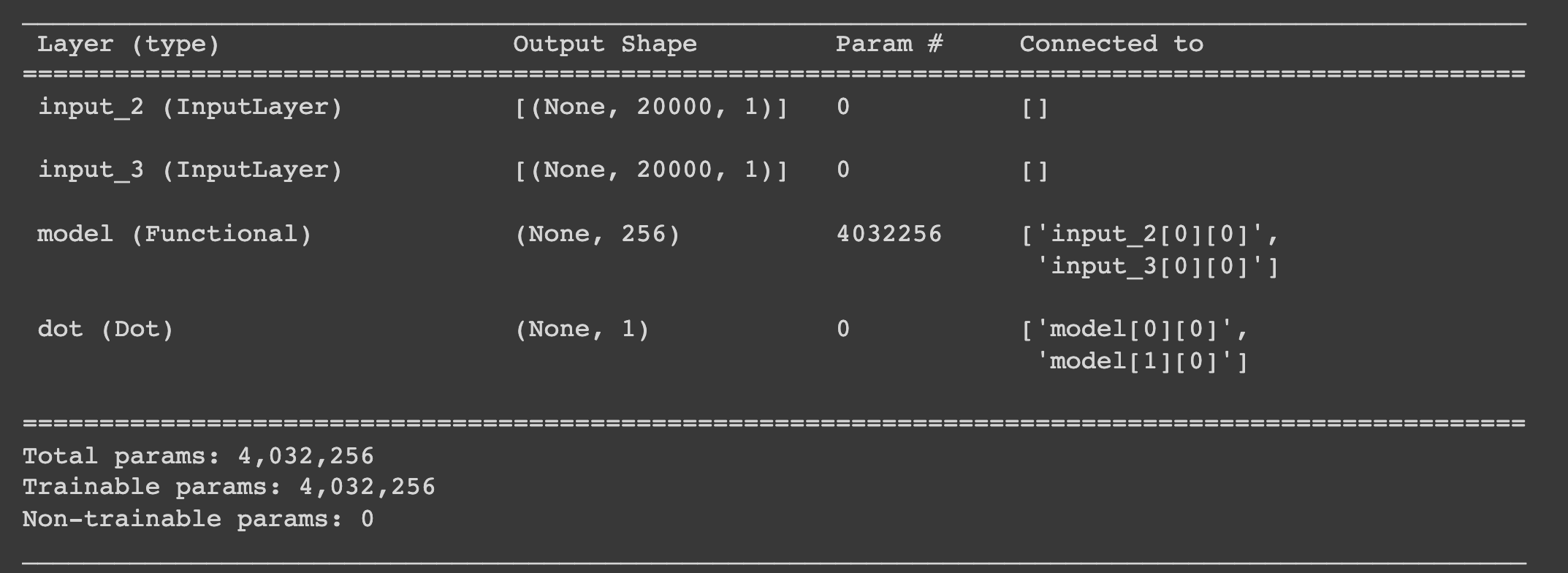}
    \caption{Siamese Network Model}
    \label{fig:model_details}
\end{figure*}

\begin{table*}[]
  \centering
  \caption{Total error of the first PASCAL classifying heart sounds challenge found by our methodology and by other approaches}
  \begin{tabular}{|l|l|l|l|l|l|}
    \hline
    \multicolumn{1}{|c|}{\textbf{Paper}}    & \multicolumn{1}{c|}{\textbf{Total Error}} & \multicolumn{1}{c|}{\textbf{Total Error in Seconds}} \\ \hline
    ISEP/IPP Portugal        & 4219488 & 95.68                  
    \\ \hline
    CS UCL &  3394377 &  76.97
    \\ \hline
    
    SLAC Stanford & 1243620 &  28.2                            
    \\ \hline
    UPD DCS Philippines & 3012912 &  68.32 
\\ \hline
signals processing approach\cite{SignalProcessing} &857304   &19.44
\\ \hline
Our Method &335362.0 &7.60 
\\ \hline
  \end{tabular}
  \label{table}
\end{table*}

\textbf{Onset Corrections and Backtracking Algorithm:} Heart State sequence is determined, S1 is followed by S2 and vice versa. From the above method, we get a State sequence and some corrections might be needed, we propose this algorithm. Firstly, we scan through the State sequence and find the non-alternating pairs and try to fit in onsets to make the sequence proper. That is, if S1 is found after S1, we try to fit in an S2 onset in between them. Let the first S1 be called S11 and the second is S12 and their corresponding times are $t_1$ and $t_2$, then the time of the new S2 is 

\begin{equation}
    t_{s2} = t_1 + \frac{\tau_1}{\tau_1 + \tau_2}*(t_2-t_1)
\end{equation}

And if we find an improper pair of S2 and S2, we insert an S1 in between. Let the first S2 be S21 and the second S2 be S22 and their corresponding times be $t_1$ and $t_2$, then the time of the new S1 is

\begin{equation}
    t_{s1} = t_1 + \frac{\tau_2}{\tau_1 + \tau_2}*(t_2-t_1)
\end{equation}

Backtracking: In the above method, we find the peaks of S1 and S2, not the beginning of S1 and S2. To get the beginning we backtrack in time to find the corresponding local minimum in the energy envelope.

\textbf{Heart Rate finding:} From the above State sequence we can extrapolate the heart beats per minute. We take S1, S2 pair duration and average it on all the available S1, S2 pairs and then using that we find out the heart rate per minute:

\begin{equation}
    Heart Rate = \frac{1}{avg\_\Delta t}*60
\end{equation}

\subsection{\textbf{Using Siamese Network to Identify S1 and S2}}
Another method to find the S1 and S2 sounds from the detected onsets is using Deep Neural Network. Given part of an audio signal between two onsets, we find whether it belongs to S1 or S2. For this, we can interpret the audio signal as an image. We can do the same for all the labelled S1's from the PASCAL dataset and interpret them as images. Now we can define the problem statement as finding whether a given audio-onset-image belongs to S1 class of images or not using all the S1 class images of the PASCAL dataset. For this problem statement, as the available data from the PASCAL dataset \cite{PASCAL} is very limited, we can use Siamese CNN Networks, as they are excellent at finding similarity between images using sparse data.
The advantage of using Siamese network for finding audio similarity is, it learns meaningful features of the signal directly from the data, and thereby it is robust to different acoustic settings. To the authors' knowledge, this is the first method to use Siamese CNN network for S1 and S2 heart sound segmentation. The Siamese Network has two inputs for the two CNN networks and these CNNs have the same structure and share the same weights. In the supervised learning setting the Siamese network tends to maximize the negative pair spatial distance and minimize the spatial distance between the positive pairs. Here, the positive pair refers to the samples belonging to the same class and the negative samples refer to the pairs belonging to different classes. The networks which make up a Siamese network is called the base network. In this case of Siamese CNN network, the base network is a Convolutional Neural Network. We feed the base CNN an event that is given onsets $O_1,O_2..O_n$ an event is $(O_t , O_{t+1})$ we can treat this as a one-dimensional image.

Firstly, we use an input layer which takes in the 1D audio image as input. Then we have a convolution layer, then we have a max pool layer, then from the output we flatten it and from that, we add a dense layer. We then normalized the output from the dense layer as it will be a feature for the audio image, we do the same for the two networks and the resultant output feature from both we perform a dot product. As the features are normalized, the resultant of the dot product will be in the range of 0 to 1, so we can use this to assign to a target label. The model summary is present in figure 6.

\textbf{Dataset Preparation:} For creating the dataset we have used the PASCAL Challenge one dataset \cite{PASCAL} which has the .wav file of the signals and the exact position of the S1 and S2 heart sounds. We use these S1 sounds as class one and S2 sounds as class two. We extract the S1 sounds and S2 sounds part of the signal from the original signal and make two classes, class1 and class2. Taking class1 as our primary focus, we create a positive pair samples set of all possible pairs. Then we create a negative pair sample set using class1 as our primary and class2 as secondary members of the pair. This creates a huge positive set and negative set despite having a very limited heart signal samples in PASCAL dataset. 

\section{PERFORMANCE EVALUATION}
In this section, we present the performance of using the DP algorithm and the Siamese network. Given the audio signals, we first perform bandpass filtering, then onset detection, and then we use our DP algorithm. The performance of this method is better than the 3 finalists of the PASCAL Classifying Heart Sounds Challenge (ISEP/IPP Portugal, CS UCL and SLAC Stanford) and \cite{SignalProcessing}.

\textbf{Siamese network:} From the detected onsets we can utilize Siamese network as well to find whether a given interval between two onsets belong to S1 or S2. The dataset we used for training this Siamese network is randomly chosen ten samples from the PASCAL dataset. We found that considering the problem as two class (S1 and S2) classification we observed low accuracy, so we redefined the problem as single class (S1) classification problem. We use Siamese network to find whether an interval between two consecutive onsets belong to S1 or not. After preparing the dataset we trained it and the training accuracy was above 95 percent. Using remaining samples from the PASCAL we created a testing data and the test accuracy was above 90 percent.

\bibliographystyle{plain}
\bibliography{bibfile}

\end{document}